\documentclass[twocolumn,pra,aps, showpacs]{revtex4-1} %
\usepackage[T1]{fontenc}                     %
\usepackage{textcomp}                        %
\usepackage{amsmath}            %
\usepackage{amssymb}            %
\usepackage{graphicx}           %
\makeatletter                   %
\usepackage{physics}

\makeatother

\begin{document}

\title{Turnplate of quantum state}

\author{Yang Liu}

\affiliation{Beijing National Laboratory for Condensed Matter Physics,
  and Institute of Physics, Chinese Academy of Sciences, Beijing
  100190, China}

\author{D. L. Zhou}

\email{zhoudl72@iphy.ac.cn}

\affiliation{Beijing National Laboratory for Condensed Matter Physics,
  and Institute of Physics, Chinese Academy of Sciences, Beijing
  100190, China}

\begin{abstract}
  Time-reversal symmetry breaking can enhance or suppress the
  probability of success for quantum state transfer (QST), and
  remarkably it can be used to implement the directional QST.\@ In
  this paper we study the QST on a ring with time-reversal asymmetry.
  We show that the system will behave as a quantum state turnplate
  under some proper parameters, which may serve as time controlled
  quantum routers in complex quantum networks. We propose to to
  realize the quantum state turnplate in the coupled resonator optical
  waveguide by controlling the coupling strength and the phase.
\end{abstract}


\maketitle

\section{Introduction}
\label{sec:intro}

Quantum state transfer (QST) is one of the basic tasks in quantum
information process. In the past decade QST has been studied
intensively. Several schemes are proposed to achieve it by different
channels, i.e., spin chains~\cite{PhysRevLett.91.207901,
  PhysRevLett.92.187902, PhysRevLett.106.040505,Yao2012,
  PhysRevA.71.022301}, polarized photons in the optical
fiber~\cite{PhysRevLett.78.3221, Ritter2012}, coupled-cavity array and
so on. Based on cavity quantum electrodynamics a scheme is to transfer
the state of a qubit from a cavity-atom system to another one through
an optical fiber connecting the two cavities. Using the spin chain as
a channel many schemes are reported, such as, QST along a one
dimensional unmodulated spin chain~\cite{PhysRevLett.91.207901},
perfect QST achieved by modulating the coupling
strength~\cite{PhysRevA.72.034303, PhysRevA.76.052328,
  PhysRevA.71.022301, 1367-2630-12-2-025019, PhysRevA.78.022325,
  PhysRevLett.106.040505, PhysRevA.85.022312}, QST without
initialization~\cite{PhysRevLett.101.230502, PhysRevA.79.054304},
optimizing basis~\cite{Bayat2011,0295-5075-102-5-50003} and
generalizing to the high spin QST~\cite{Bayat2007,Qin2013,
  PhysRevA.89.062302}. Other schemes, such as, transferring
single-mode photon state through a coupled-cavity array, are also
reported. Recently time-reversal symmetry breaking is introduced to
study the QST~\cite{Zimboras2013, Lu2014}, where the time-reversal
symmetry breaking can enhance or suppress the probability of the QST
and make the QST directional bias. In this sense time asymmetry is a
new resource for exploring the QST.\@

In Refs.~\cite{hafezi_robust_2011, hafezi_imaging_2013}, it is shown
that a synthetic magnetic field can be introduced for photons by
differential optical paths in system of the coupled resonator optical
waveguides (CROW). In this paper we consider the QST along a ring
consisting of coupled cavities or coupled resonator optical waveguides
with time reversal asymmetry. Because of the time reversal symmetry
breaking we hope that the quantum state transfer along the ring one by
one periodically like a turnplate of quantum states. The quantum
turnplate will be useful in building complex quantum networks where it
acts as a quantum router. In the following of the paper we show that
a CROW ring will behave as quantum state turnplate under some proper
parameters.

This article is organized as follows. First we give the physical model
and make a general analysis. Then, we analyze the dynamic requirement for
the quantum state turnplate in a single excitation model, and give the
energy spectrum and symmetry matching condition for quantum state
turnplate. In Sec.~\ref{sec:structure-spectrum}, we study the
spectrum of the system with the $c_n$ symmetry. Then, we discuss the
effective Hamiltonian of the system using perturbation method. We come
back to the CROW system in Sec.~\ref{sec:quant-turnpl-crow}.
Finally a summary is given.

\section{Physical model}
\label{sec:physical-model}

\begin{figure}[htb]
  \includegraphics[width=4cm]{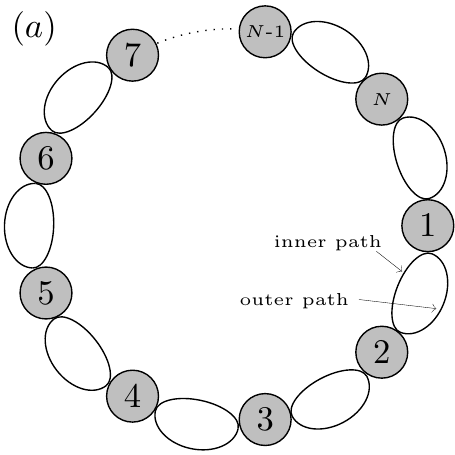}
  \includegraphics[width=4cm]{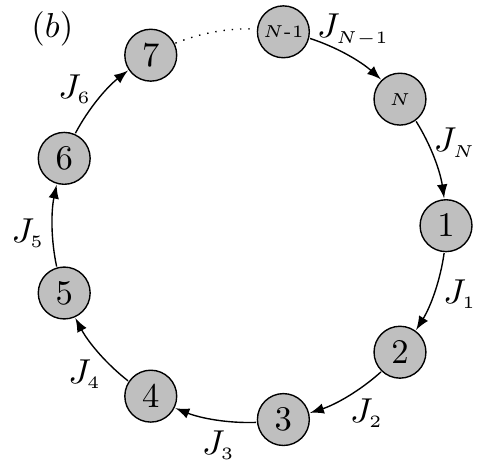}
  \caption{(a) The CROW ring. The circles mean the site resonators,
    and the link optical waveguides link the resonators as a
    ring. Every waveguide between two site resonators has two paths,
    the outer path and the inner path. The two paths have different
    length which induce the complex coupling strength $J$. (b) Single
    excitation graph. The circles correspond to the resonators in CROW
    system. The arrow with labels $J_l$ represents the matrix element
    $J_l|l\rangle \langle l+1|$.}\label{fig:1}
\end{figure}

In the CROW the synthetic magnetic field can be introduced by
differential optical paths~\cite{hafezi_robust_2011,
  hafezi_imaging_2013}. We consider the CROW system in a ring
configuration as shown in Fig.~\ref{fig:1}(a).  The Hamiltonian of
the ring is
\begin{equation}
  H_R = \sum_{l=1}^{N} (J_l \hat{a}_l \hat{a}^{\dagger}_{l+1} +J_l^{\ast} \hat{a}^{\dagger}_l \hat{a}_{l+1}),
\end{equation}
where $J_l$ is the coupling strength of between the sites $l-1$ and
$l$, and $N+1$ is interpreted as $1$. $\hat{a}_l$
($\hat{a}_l^\dagger$) is the annihilation (creation) operator. From
Ref.~\cite{liu_transfer_2015} we know that the condition for
transferring any single mode photon state from node $l$ to node $l'$
is
\begin{equation} 
  \hat{a}_{l'} (\tau) = \hat{a}_l,
\end{equation} 
where $\hat{a}_{l'}(\tau) = U^{\dagger}(t) \hat{a}_{l'} U(t)$ with
$U(t)$ being the time evolution operator. It can be easily verified by
noting that the expect value of any operator in the $l'$-th node at
time $\tau$ is equal to that of the operator in the $l$-th node in the
initial state.

Using the Heisenberg equation 
\begin{equation}
  \frac{d \hat{a}_l(t)}{dt} = i \left[ H_R, \hat{a}_l(t) \right]
\end{equation}
and noting that $\hat{a}_l(t)$ can be expressed on the operator bases as
\begin{equation*}
  \hat{a}_l(t) = \sum_{k=1}^{N} \alpha_k(t) \hat{a}_k ,
\end{equation*}
the evolution of the operator $\hat{a}_l(t)$ can be written as 
\begin{equation}
  i\frac{dA}{dt} = -H(J) A, 
  \label{eq:evolution_equation}
\end{equation}
where $A={[\alpha_{1}(t), \, \alpha_{2}(t), \,\ldots,\alpha_{N}]}^{T}$
with $T$ being the transpose operation, and
\begin{equation}
  H(J)= 
  \begin{bmatrix}
    0 & J_{1} & 0 & \cdots & J_{N}^{*} \\
    J_{1}^{*} & 0 & J_{2} & \cdots & 0 \\
    0 & J_{2}^{*} & 0 & \cdots & 0 \\
    \vdots & \vdots & \vdots & \ddots & \vdots \\
    J_{N} & 0 & 0 & \cdots & 0
  \end{bmatrix},
  \label{eq:1}
\end{equation}

The initial condition is $A(0)={[0,\,0,\,\ldots, 1, \ldots, 0]}^{T}$,
where $1$ is the $l$-th element.  With the above initial condition,
Eq.~\eqref{eq:evolution_equation} describes the single excitation
evolution in the ring with coupling strength $\{ -J_i \}$.

So the transfer of any single mode photon state in the CROW ring has
the same physical picture as the single excitation model, and we do
not need to initialize the state of other sites except the input one.

\section{Single excitation ring}
\label{sec:single-excit-model}

Firstly, we consider the single excitation model. To sketch our
central idea, we consider the system depicted by a graph in
Fig.~\ref{fig:1}(b) consisting of $N$ sites as a ring, with the
Hamiltonian $H(J)$ having the form given in Eq.~\eqref{eq:1}, where
$J$ is the the set $\{J_{l}\}$. Let us denote $\Phi$ as the set
$\{\phi_l\}$, where $\phi_l$ is the complex phase of $J_l$. From
Ref.~\cite{Lu2014} we know that the phase $\Phi$ can give rise to the
time-reversal asymmetry if the graph is non-bipartite graph ($N$ is
odd). In this article we concentrate on the cases where $N$ is odd.

Through local unitary operator $U^{L}$, the Hamiltonian can be
transformed to
\begin{equation*}
  H(J')=U^{L}H(J)U^{L \dagger},
\end{equation*}
where $\sum_{l}\phi'_{l}=\sum_{l}\phi_{l}$ and $|J'_l|=|J_l|$. In
other words, only the sum of phases $\sum_l\phi_l$ is relative to the
properties of QST.\@ So we can
choose a proper operator $U^{L}$ to make all the phases equal,
$\phi_{l}'=\frac{\sum_{l}\phi_{l}}{N}$, with the QST properties in the
time evolution unchanged.

Now we consider the question: In what condition would the system
behave like a turnplate of quantum state? Firstly, we require that the
system have the symmetry of the cyclic group, $c_{n}$, for the
turnplate having $n$ ($n\leq N$) scales on it, i.e. $N/n$ is an
integer. The operators of the $c_{n}$ group elements can be expressed
as
\begin{equation*} 
  \mathcal{T}_n,\, {(\mathcal{T}_n)}^2,\, \ldots,\,
  {(\mathcal{T}_n)}^{n-1},\, 1
\end{equation*}
where
\begin{equation*}
  \mathcal{T}_{n}=e^{iL_{n}\frac{2\pi}{n}}
\end{equation*} 
and $L_{n}$ is a Hermitian operator. From ${(\mathcal{T}_n)}^n = 1$, we
know that $L_n$ has $n$ integer eigenvalues,
$l\in\{\lfloor-\frac{n-1}{2}\rfloor,\,\lfloor-\frac{n-1}{2}\rfloor+1,\ldots,\,\lfloor\frac{n-1}{2}\rfloor\}$
，where $\lfloor x\rfloor$is the largest integer not greater than
$x$. Here we only consider the case where $n$ is odd. For the
Hamiltonian we have the relation $[H,\, L_{n}]=0$. Let the eigenstate
of the system be $\left|\psi_{l,m}\right>$, that is
\begin{equation*} 
  H|\psi_{l,m}\rangle=E_{l,m}|\psi_{l,m}\rangle, 
\end{equation*} 
and 
\begin{equation*} 
  L_n |\psi_{l,m}\rangle = l |\psi_{l,m}\rangle.
\end{equation*}

Now we prove that the system will be a turnplate of quantum states with
$n$ scales if the eigenvalues $E_{l,m}$ match the symmetry $c_n$ in
the following way:
\begin{equation}
  E_{l,m} = \left(\frac{l}{n}+Z_{m}\right)\epsilon+\epsilon_{0}, \label{eq:2}
\end{equation}
 where $Z_{m}$ is an integer, and $\epsilon_0$ correspond to the phase
 that is not an observable in physics.

 Let the initial state of the system is $|\psi_{0}\rangle$. It can be
 easily proved that at time $\tau=\frac{2\pi}{\epsilon}$,
\begin{equation}
|\psi(\tau)\rangle= e^{i \pi \epsilon_0
   /\epsilon}\mathcal{T}_{n}|\psi_{0}\rangle, \label{eq:3}
\end{equation}
 meaning that quantum state of the system turn one scale every time
 interval $\tau$. The energy and symmetry matching condition can be seen as the
 generalization of the energy and parity matching condition mentioned
 in Ref.~\cite{PhysRevA.71.032309}, where the parity matching
 condition corresponds to the case $n=2$ and $\epsilon=2E_0$.

\begin{figure}[htb]
\includegraphics{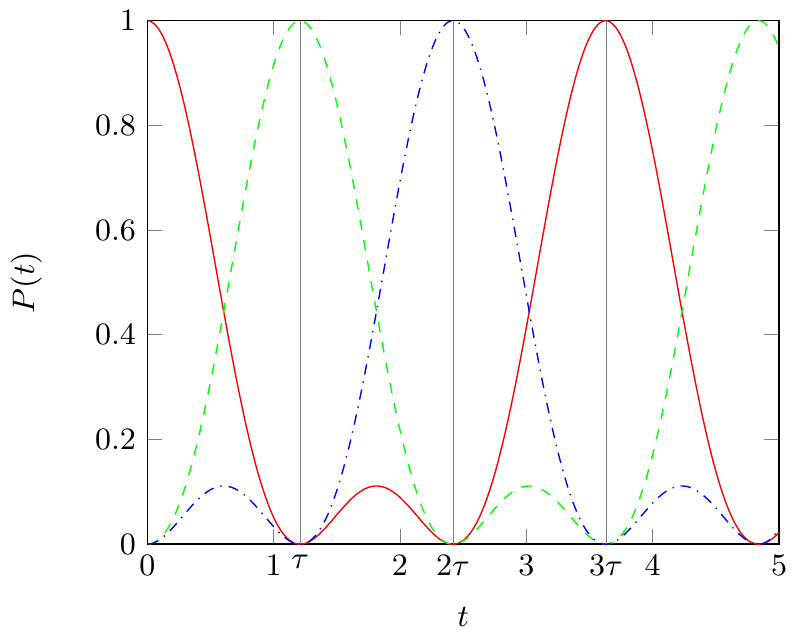}
\caption{\label{fig:triangle} (Color online) Numerical
  stimulation of the time evolution of the system consisting of 3
  sites with the parameters $J_1 = J_2 = J_3 = e^{i\frac{\pi}{6}}$ and
  the initial state $|100\rangle$. $P(t)$ is the probability of the
  wave function. The solid line, dashed line and dashdotted line
  describe the probability at site 1, site 2 and site 3
  respectively. At time $\tau$ the excitation transfers from site 1 to
  site 2 and it transfers to site 3 after a next time interval
  $\tau$. The system acts like a turnplate of the excitation.}
\end{figure}

 Here we look at the particular case $N=n=3$. The relation
 $[H,\,\mathcal{T}_{3}]=0$ requires that
 $J_{1}=J_{2}=J_{3}$. $l\in\{0,\,\pm1\}$ and using the property of the
 $c_{3}$ group we known $H$ and $L_{n}$ have the same eigenvectors
 $|\phi_{l}\rangle=\frac{1}{\sqrt{3}}{[1,\,\omega_{3}^{l},\,\omega_{3}^{2l}]}^{T}$
 with $\omega_{3}=e^{i\frac{2\pi}{3}}$. When
 $J_1 = J_2 = J_3 = e^{i\frac{\pi}{6}}$, that is the total phase is
 $\pi/2$, the eigenvalues are $E_{l=0}=\sqrt{3}$, $E_{l=1}=-\sqrt{3}$,
 $E_{l=-1}=0$. So we get $\epsilon_{0}=\frac{\epsilon}{3} = \sqrt{3}$
 and the time interval is $\tau = \frac{2\pi}{3\sqrt{3}}$. We
 numerically stimulate the time evolution of this case and show the
 probability of the wave function in the Fig~\ref{fig:triangle}.

Given the single excitation condition, it was proved that there is the
pretty good state transfer between any two sites of a uniform ring
with total phase as $\pi/2+k\pi$, if the number of the site of the
ring is prime, in Ref.~\cite{cameron_universal_2014}. It indicates
that the energy spectrum and symmetry matching condition,
Eq.~\eqref{eq:2} is approximately satisfied when the length of the
ring is prime.

\section{structure of the spectrum}
\label{sec:structure-spectrum}

In this section we analyze the energy spectrum of the system described by the
Hamiltonian in Eq.~\eqref{eq:1}. We start with the definitions of some
notations. We denote the characteristic polynomial as $A_{N}$, that is
$A_{N}=\det(H_{N}-\lambda)$. And denote $B_{N}$ as
$B_{N}=\det(H_{N}(J_{N}=0)-\lambda)$, where $H_{N}(J_{N}=0)$
represents the Hamiltonian where the coupling between first site and
the last site is zero, i.e., the chain is an open one.  From the
Ref.~\cite{PhysRevA.89.062331} we know that,
\begin{equation*}
  B_{N}=\begin{cases}
    \lambda g(\lambda^{2}) & \text{ if }N\text{ is odd,}\\
    g(\lambda^{2}) & \text{ if \ensuremath{N}is even, }
  \end{cases}
\end{equation*}
where $g\left(x\right)$ is an arbitrary function of $x$.  So the
determinate of $A_N$ with odd $N$ is
\begin{align*}
  \det A_{N} & =-\lambda\det B_{N-1}-J_{1}^{2}\det B_{N-2}\\
             & \ \ \ +\prod_{l=1}^{N}J_{l}e^{i\phi}+\prod_{l=1}^{N}J_{l}e^{-i\phi}-J_{N}^{2}\det B_{N-2}\\
             & =\lambda f(\lambda^{2})+\lambda g(\lambda^{2})+2\prod_{l}J_{l}\cos\phi\\
             & =\lambda F(\lambda^{2})+\prod_{l}J_{l}2\cos\phi，
\end{align*}
where $\phi$ is the total phase. When $\phi=\frac{\pi}{2}+k\pi$,
$k=0,\,\pm1,\,\pm2,\,\cdots$, the spectrum has the structure
$\{0,\,\pm E_{l}\}$ that is the spectrum is symmetric around $0$. Let
us consider the system that has the $C_{n}$ symmetry and contains
$N=n\times p$ sites. The eigenvalues of the operator $L_{n}$ is
$0,\,\pm1,\,\ldots,\,\pm\frac{n-1}{2}$, which we label as $l$, and
every eigenvalue has $p$-fold degeneracy. We can easily write out the
eigenvector of $L_n$,
\begin{align*}
  |l,\, i\rangle & =\frac{1}{\sqrt{n}}\left(|i\rangle+\omega_{n}^{l}
                   |i+p\rangle + \omega_{n}^{2l}|i+2p\rangle\right.\\
                 & \left.+\cdots+\omega_{n}^{(n-1)l}|i+(n-1)p\rangle\right).
\end{align*}
Let $P_{l}=\sum_{i}|l,i\rangle\langle l,i|$. From the relation $[H,\,
L]=0$, we know that
\begin{alignat*}{2}
  P_{l}HP_{l'} & =0 & \ \ \ for\text{ }l\neq l'.
\end{alignat*} 
This means that the Hamiltonian is block diagonalized under the bases
$\{ |l,\, i\rangle \}$. Because the system has the $c_{n}$ symmetry,
we have the relation $J_{i}=J_{i+p}$. So there are only $p$
parameters, $J_{1},\, J_{2},\,\ldots,\, J_{p}$. We can always let
$J_{1}=1$ and other parameters be the ratio to $J_{1}$. Then the
property of the system don't change up to the time scales. So there
are $p-1$ parameters that we need to consider.

\begin{figure}[htb]
\centering

\includegraphics{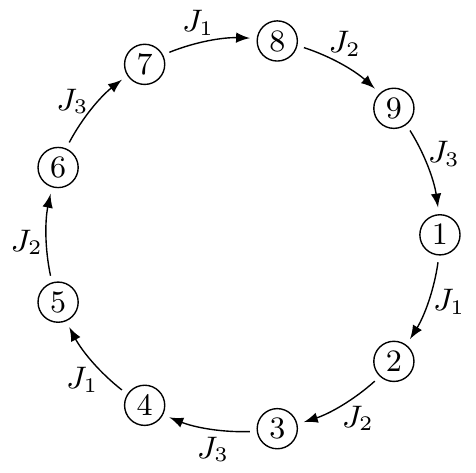}

\includegraphics{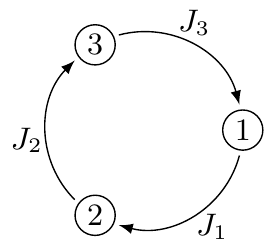}\includegraphics{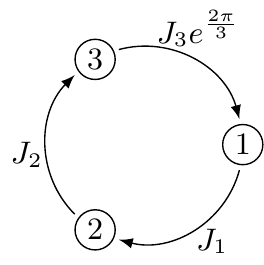}\includegraphics{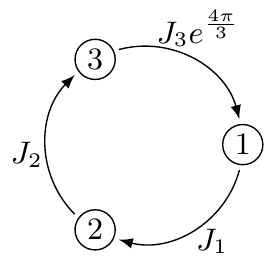}

\protect\caption{The ring with $9$ nodes and $c_3$ symmetry. Under
  the bases $|l,i\rangle$ the Hamiltonian is block diagonalized
  consisting $3$ blocks which is implied by $c_3$. In every block
  the Hamiltonian represents an ring with length $3$ and the coupling
  strength is the same as the original ring in the site bases with
  the total phase being $\phi/3$, $\phi/3 + 2\pi/3$ and
  $\phi/3+4\pi/3$ where $\phi$ is the total phase of the original ring.\label{fig:3}}
\end{figure}

\begin{figure}[htb]
  \centering
  \includegraphics{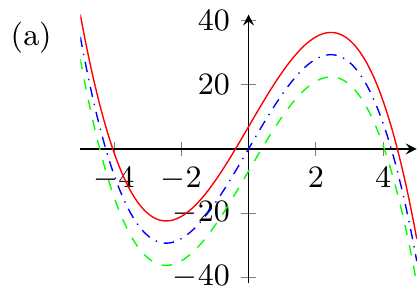} 
%  \resizebox{1cm}{3cm}{\includegraphics{figure/energy_lever5}}
  \includegraphics{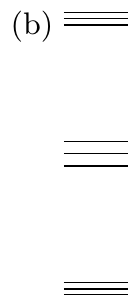}

  \caption{\label{fig:4} Spectrum structure of the Hamiltonian $H_9$
    with length $N=9$ and $C_3$ symmetry. The coupling strength is
    $J_1=J_3=e^{\frac{\pi}{18}}$, $J_2=4J_1$. So the total phase is
    $\phi=\pi/2$. (a) Pictures of the characteristic polynomials of
    three equivalent Hamiltonians got from the Hamiltonian $H_9$. The
    dashdotted line is the curve-$0$ corresponding to $l=-1$, and the
    other two lines can be got from curve-$0$ by translated along the
    vertical axis. (b) Spectrum of the Hamiltonian $H_9$. Every energy
    elver correspond to one cross point of the curve and the
    horizontal axis in (a).}

\end{figure}

Using the bra ket form of the Hamiltonian
\begin{align}
  H & =J_{N}|N\rangle\langle1|+J_{N}^{*}|1\rangle\langle N| \nonumber\\
    & \ \ \ +\sum_{k=1}^{N-1}\left(J_{k}|k\rangle\langle
      k+1|+J_{k}^{*}|k+1\rangle\langle k|\right), \label{eq:H-all}
\end{align}  
and acting the projector $P_l$ on both sides of Eq.~\eqref{eq:H-all}
we can directly give
\begin{align}
  P_{l}HP_{l} &
                =\omega_{n}^{l}J_{p}|p\rangle\langle1| +
                \omega_{n}^{l*}J_{p}^{*}|1\rangle\langle p|
                \nonumber\\ 
              & \ \ \ +\sum_{k=1}^{p-1}\left(J_{k}|k\rangle\langle
                k+1|+J_{k}^{*}|k+1\rangle\langle k|\right). \label{eq:H-block}
\end{align}
Comparing Eq.~\eqref{eq:H-all} and Eq.~\eqref{eq:H-block} we get that
in the every block the Hamiltonian is equivalent to the Hamiltonian of
the ring with length $p$ and the moduli of the coupling strength are
not changed just with the total phase changing from $\phi /n$ to
$\phi_{l}=\frac{\phi}{n}+\frac{2l\pi}{n}$, see Fig.~\ref{fig:3}. So
the characteristic function can be written as
\begin{equation*}
\lambda F(\lambda^{2})+2\prod_{k=1}^{p}J_{k}\cos\phi_{l}=0,
\end{equation*}
where $l=-\frac{n-1}{2},\,\ldots,\,\frac{n-1}{2}$. Every function
means a curve that cross with the axis of the variable $p$ times
corresponding to the $p$ roots, see Fig.~\ref{fig:4}(a). All the
curves have the same shape. When the total phase
$\phi = \frac{\pi}{2} + k \pi$, the curve corresponding to
$l=-\frac{n+1}{4}$ (or $l=\frac{n-1}{4}$) crosses the original point and
we call it curve-$0$. Other curves can be got from the curve-$0$ by
translate $2\prod_{k}J_{k}\cos\phi_{l}$ along the vertical axis. So if
$\prod_{k=1}^{p}J_{k}$ are little enough the spectrum of the
Hamiltonian has the shape indicated in Fig.~\ref{fig:4}(b), that is,
the spectrum consists of separated groups.

\section{Effective Hamiltonian}
\label{sec:effect-hamilt}

Now we introduce the approximate method based on the spectrum
structure. To discuss concretely, we consider the
system with nine sites ($N=9$) and the $c_{3}$ symmetry. Its configure is
shown in Fig.~\ref{fig:3}. The eigensystem of the Hamiltonian $H$ is
equivalent to the Hamiltonian $H'$ with
\begin{alignat*}{1}
  &J'_{k}=|J_{k}|,\text{ for }k\neq1\\
  & J'_{1}=|J_{1}|e^{i\phi},
\end{alignat*}
where $\phi = \sum_k \phi_k$. So we consider the Hamiltonian
$H'$. Let $J_{2}'\gg J_{1,3}'$ and write the Hamiltonian $H'$ into two
terms, $H'=H'_{0}+V$. Given that the $H'$ is represented in the site
basis $|i\rangle$, $H'_{0}$ consists of the terms containing $J_{2}'$
, and $V$ consists the terms containing $J'_{1,3}$. $V$ is the
perturbation compared with $H_{0}'$.

The eigenvalues of the Hamiltonian $H'_{0}$ are $\alpha\in\{0,\,\pm J_{2}^{'}\}$
and every energy level has three-fold degeneracy. So the energy levels
are separated into three groups (manifold) corresponding to three $\alpha$s.
Using $i$ label the different bases we denote the three manifold
as $|i,\,\alpha\rangle$. In the manifold with $\alpha=0$ the three
bases are 
\begin{equation*}
|1\rangle,\,|4\rangle,\,|7\rangle.
\end{equation*}
 And the manifolds with $\alpha=\pm|J_{2}|$ are spanned by the bases
\begin{equation*}
|X_{23}^{+}\rangle,\,|X_{56}^{+}\rangle,\,|X_{89}^{+}\rangle
\end{equation*}
 and 
\begin{equation*}
|X_{23}^{-}\rangle,\,|X_{56}^{-}\rangle,\,|X_{89}^{-}\rangle
\end{equation*}
 respectively, where 
\begin{equation*}
  |X_{ij}^{+}\rangle=\frac{|0\rangle_{i}+|1\rangle_{j}}{\sqrt{2}},\text{
    and
  }|X_{ij}^{-}\rangle=\frac{|0\rangle_{i}-|1\rangle_{j}}{\sqrt{2}}.
\end{equation*}

\begin{figure}[htb]
\includegraphics{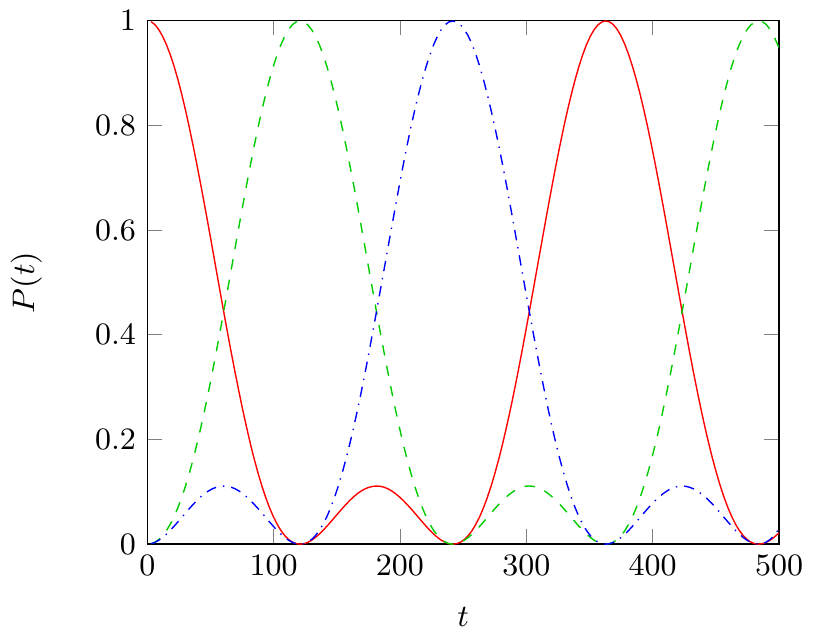}
\protect\caption{\label{fig:5} (Color online) Numerically simulate the
  time evolution of the system with 9 sites. $J_1=J_3=1$, $J_2=100$
  and $\phi=-\frac{\pi}{2}$. The initial state is
  $|\psi(0)\rangle=|100000000\rangle$ with $|1\rangle$ as input state.
  Solid line presents the fidelity of the state transfer for the state
  of the first site at different time. The dashed line and the
  dashdotted line corresponding to the fidelity for the seventh site
  and fourth site respectively. }

\end{figure}

We take $V$ as perturbation then compute the effective Hamiltonian
in the manifold $\alpha=0$. And the effective Hamiltonian is

\begin{equation*}
  H_{\rm eff}^{\alpha=0}=-\begin{bmatrix}0 & ge^{i\phi} & g\\
    ge^{-i\phi} & 0 & g\\
    g & g & 0
  \end{bmatrix},
\end{equation*}
where $g=\frac{J_{1}J_{3}}{J_{2}}$. $H_{\rm{eff}}^{\alpha=0}$ is
identical to the representation of the Hamiltonian of the ring
consisting of three nodes with uniform coupling strength $g$ and total
phase $\phi$. From the analysis in the
Sec.~\ref{sec:single-excit-model} we know that when
$\phi=\frac{\pi}{2}+k\pi$ the system is a turnplate of quantum state
with three scales and the time interval of transfer state from one
node to next is $\frac{2\pi}{3g\sqrt{3}}$. We numerically simulate the
time evolution of the system with parameters $J_1=J_3=1$, $J_2=100$
and $\phi=-\frac{\pi}{2}$ from the initial state
$|\psi(0)\rangle=|1\rangle$ in Fig.~\ref{fig:5}. At time $\tau=120.92$
the excitation transfer from site $1$ to site $4$ and after the same
time interval it transfer to site $7$ then back to site $1$
circularly. The sites except the sites $1$, $4$ and $7$ can't be
excited.

\begin{figure}[htb]
  \includegraphics[width=8cm]{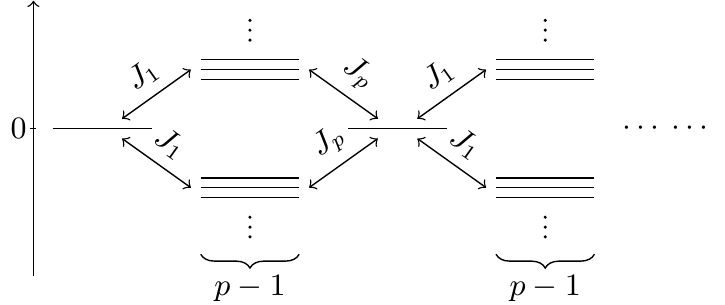}
  \caption{\label{fig:effective_lever} Spectrum of the $H_0$ which is
    the main part of the Hamiltonian. Every zero energy level denotes
    the Hilbert space of the sites labeled by number $(1+l \times p)$,
    and every grouped $p-1$ levels denote the Hilbert space of the
    sites form $(2 + l \times p)$ to $p + (l\times p)$ respectively,
    where $l=0,1,\ldots,n-1$.}
\end{figure}

The correspondence between the propriety of the system with site
number $N=n\times p$ and $N=n$, where $n$ is decided by the symmetry
of the large system, can be generalized to general case. Let
$J_{1(p)}\ll J_{l\neq 1(p)}$ then the Hamiltonian can be written as
$H=H_0 + V$, where $V$ is perturbation term consisting of the terms
containing $J_{1(p)}$, and $H_0$ is other terms. The spectrum of the
Hamiltonian $H_0$ has the form depicted in
Fig.~\ref{fig:effective_lever}. The zero energy level is $n$-fold
degeneracy with degenerate ket $|m^{(0)} \rangle$, where
$m=1+l\times p$ and $l=1,2,\ldots,n-1$. All the $|m\rangle$s span the
manifold $\mathcal{M}_{\alpha=0}$. The energies greater and less than
zero distribute two sides of the zero energy level symmetrically with
an energy gap and every energy level is $n$-fold degeneracy. From the
perturbation theory in the degenerate case we know that up to the
first order the eigenkets of Hamiltonian $H$ corresponding to the
manifold $\mathcal{M}_{\alpha=0}$ are
\begin{equation*}
|\phi \rangle =\sum_m c_m |m\rangle - \sum_{k\notin \mathcal{M} }
\frac{|k^{(0)}\rangle V_{k,m}}{E_k^{(0)}},
\end{equation*} 
where $V_{k,m}=\langle k| V |m \rangle$ and $|k^{(0)}\rangle$ is the
eigenket of $H_0$ which is not in the manifold
$\mathcal{M}_{\alpha=0}$. So when $V_{k,m}$ is much less than $E_k^{(0)}$
(the gap on the zero energy level), the manifold
$\mathcal{M}_{\alpha=0}$ is close, i.e., if the initial state
is $|m\rangle$ then the system is govern by the effective Hamiltonian
$H_{eff}^{\alpha=0}$ which represents a Hamiltonian of the $n$-site cycle.
So the system with $N=n\times p$ and $c_n$ symmetry can be reduced to
the system with $N=n$ and with $c_n$ symmetry. 

\section{Quantum turnplate on the CROW ring}
\label{sec:quant-turnpl-crow}
Now we come back to the physical system, the CROW ring. For the ring
containing three resonators, they can be connected by three identical
connecting waveguides, which contribute the same coupling strength
$|J_l|$ and phase $\phi_l$. In order to make a quantum turnplate we
just need to modify the optical path to make total phase
$\phi= \pi/2 + k \pi $. Initially we input the photonic state to
the node 1. Then we will see that the photonic state will transfer
from node 1 to node 2, node 3 and back to node 1 cyclicly with perfect
fidelity every time interval $\tau$. For the ring contains
$N = 3 \times p$ resonators, which has the $c_3$ symmetry, we need to
modify the coupling strength between resonators and connecting
waveguides to satisfy the condition $J_{1(p)}\ll J_{l\neq 1(p)}$ and
change the optical path to make the total phase be
$\phi = \pi/2 + k\pi$. Then photonic states can be transfer among the
site $1$, $p+1$, and $2p+1$ cyclicly with high fidelity. 

\begin{figure}[htb]
  \includegraphics{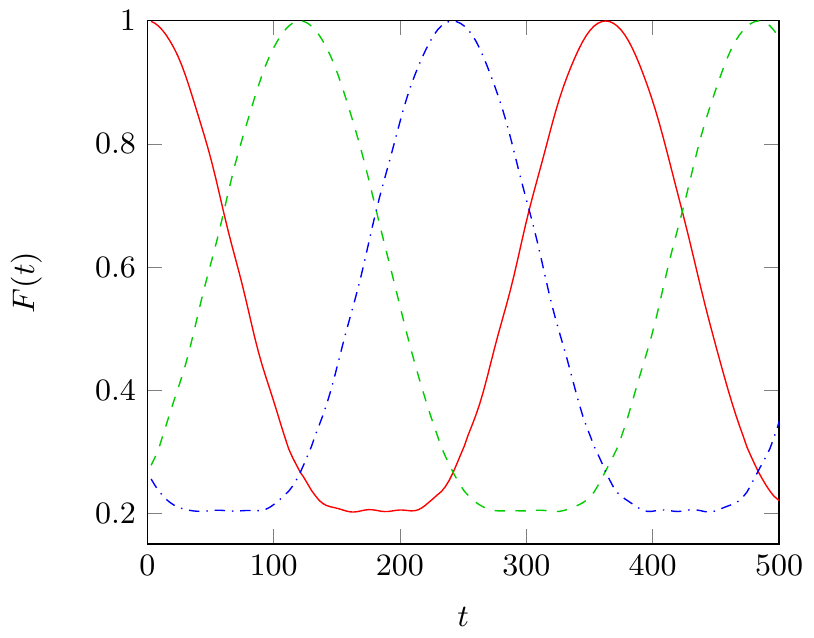}
  \protect\caption{\label{fig:6} (Color online) Numerically simulate
    the time evolution of the fidelity of the CROW system with 9
    resonators. $J_1=J_3=1$, $J_2=100$ and $\phi=\frac{\pi}{2}$. The
    initial state is
    $\frac{1}{\sqrt{28}} (|0\rangle + 2|1\rangle +3|2\rangle)
    (|1000000\rangle + |1100000\rangle)$
    with $\frac{1}{\sqrt{14}} (|0\rangle + 2|1\rangle +3|2\rangle)$ as
    the input state.  Solid line represents the fidelity of the state
    transfer for the state of the first site at different time. The
    dashed line and the dashdotted line correspond to the fidelity
    for the fourth site and seventh site respectively. }
\end{figure}

We simulate the QST along the CROW ring consisting 9 resonators with
$c_3$ symmetry. The parameters are $|J_1| =|J_3| = 1$, $|J_2| = 100$
and the total phase $\phi = \pi/2$. Initially the state
$|\psi\rangle = (|0\rangle + |1\rangle + |2\rangle)/\sqrt{3}$ is
input into the first resonator. Then we observe transfer of
$|\psi\rangle$ along the ring. In Fig.~\ref{fig:6} we plot the time evolution
of the fidelity 
\[
F(t) = \langle \psi| \rho_i(t) |\psi \rangle
\]
of sites $1$, $4$ and $7$ and it behaves as a quantum state
turnplate. $\rho_i$ means the reduced density matrix of site $i$.

\section{Discussion and summary}
\label{sec:discussion-summary}
Using the similar method we used to get the Eq.~\eqref{eq:3} we have
the equation
\begin{equation}
\hat{a}_l(\tau) = e^{i \pi \epsilon_0 /\epsilon} \hat{a}_{l-p},\label{eq:4}
\end{equation} 
for the annihilation operator of photon in the CROW ring. So the basis
for the quantum state in different site should be identified. For
example in the ring with $9$ sites the bases for site $1$ and site $4$
 should be $\{|0\rangle, |1\rangle, |2\rangle,\ldots\}$ and
 $\{|0\rangle, e^{i\pi/3}|1\rangle, e^{i2\pi/3}|2\rangle,\ldots\}$
respectively. Light scattered from the resonators can be imaged using a
infrared camera. Directly the quantum turnplate will be observed from
the image of the camera.

In summary, we study the QST on the ring of coupled cavities with time
reversal asymmetry. The transfer of any single mode photon state in
the CROW ring has the same physical picture as the single excitation
model, and we do not need to initialize the state of other sites
except the input one. To act as a quantum state turnplate the
eigenvalues of the equivalent single excitation model should satisfy
the matching condition Eq.~\eqref{eq:2}. And we show that when number
of site $N=3$ and the total phase $\phi = \pi/2 + k \pi$ the matching
condition is satisfied. Further, we study the structure of the
spectrum of the single excitation ring in general condition and prove
the QST equivalent between the ring consisting $n$ sites and the one
consisting $n\times p$ sites with $c_n$ symmetry. Utilizing the time
reversal asymmetry the CROW consisting $3 \times p$ resonators can
sever as a quantum turnplate without of initialization, which can also
be observed in experiments. Quantum state turnplates would be useful
to build a complex quantum network.

\bibliography{XX_information_flux}

\end{document}